
%
%
%
%
%
%
%
%
%
%
\input uiucmac.tex

\def\ln{\log}
\PHYSREV
\tolerance 2000
\nopubblock
\titlepage
\title{\bf Renormalization Group Theory and Variational Calculations
For Propagating Fronts}

\author{Lin-Yuan Chen, Nigel Goldenfeld, and Y. Oono}
\address{Department of Physics and Materials Research
Laboratory, University of
 Illinois at Urbana-Champaign, 1110 West Green Street, Urbana, Illinois
 61801-3080,
 USA. }
\smallskip
\abstract
\noindent
We study the propagation of uniformly translating fronts into a linearly
unstable state, both analytically and numerically.  We introduce a
perturbative renormalization group (RG) approach to compute the change
in the propagation speed when the fronts are perturbed by structural
modification of their governing equations.  This approach is successful
when the fronts are structurally stable, and allows us to select
uniquely the (numerical) experimentally observable propagation speed.
For convenience and completeness, the structural stability argument is
also briefly described. We point out that the solvability condition
widely used in studying dynamics of nonequilibrium systems is equivalent
to the assumption of physical renormalizability.  We also implement a
variational principle, due to Hadeler and Rothe, which provides a very
good upper bound and, in some cases, even exact results on the
propagation speeds, and which identifies the transition from \lq linear'-
to \lq nonlinear-marginal-stability' as parameters in the governing
equation are varied.

\medskip
\noindent
Pacs numbers:  03.40.Kf, 68.10.Gw, 47.20.K
\endpage

{\bf \chapter{Introduction}}

The steady state equation for a travelling wave propagating into an
unstable state does not uniquely determine the wave speed.  As an
alternative to directly solving the initial-value problem, velocity
selection principles have been sought, which identify {\it a priori} the
wave speed that would be dynamically selected from those permitted by
the steady state equation.  Such velocity selection and/or wave-number
selection problems\REFS\dee{G. Dee and J. S. Langer, \journal Phys.
Rev.  Lett.  &50&383(83).}\REFSCON\benjocab{E. Ben-Jacob, H. R. Brand,
G. Dee, L. Kramer, and J. S. Langer, \journal Physica D &14&348(85).}
\REFSCON\aronson{D. G. Aronson and H. F. Weinberger, in {\sl Partial
Differential Equations and Related Topics}, edited by J. A. Goldstein
(Springer, Heidelberg, 1975); \journal Adv. Math. &30&33(78).}
\REFSCON\nozaki{K. Nozaki and N. Bekki, \journal Phys. Rev. Lett.
&51&2171(83).}\REFSCON\collet{P. Collet and J. -P. Eckmann, \journal
Commun. Math. Phys. &107&39(86); \journal Physica A
&140&96(86).}\REFSCON\shraiman{B. Shraiman and D. Bensimon, \journal
Phys.  Scr. T &9&123(85).}\REFSCON\saarloos{W. van Saarloos, \journal
Phys. Rev. Lett.  &58&2571(87); \journal Phys. Rev. A
&37&211(88).}\REFSCON\van{W. van Saarloos, \journal Phys. Rev. A
&39&6367(89).}\REFSCON\gdee{G. Dee, \journal J. Stat.  Phys.
&39&705(85); \journal Physica D &15&295(85).}\REFSCON\dvs{G. Dee and W.
van Saarloos, \journal Phys. Rev. Lett. &60&2641(88).}\REFSCON\ahlers{
G. Ahlers and D. S. Cannel, \journal Phys. Rev. Lett.
&50&1583(83).}\REFSCON \fineberg{J. Fineberg and V. Steinberg, \journal
Phys. Rev. Lett. &58&1332(87).}\REFSCON\hanna{A. Hanna, A. Saul,
 and K. Showalter, \journal J. Am. Chem. Soc.
&104&3838(82).}\REFSCON\ben{E.  Ben-Jacob, N. D.  Goldenfeld, B. G.
Kotliar and J. S.  Langer, \journal Phys.  Rev.  Lett.  &53&2110(84);
D.  A.  Kessler, J. Koplik and H. Levine, \journal Phys.  Rev.  A
&31&1712(85).}  \refmark{\dee-}\refend arise in a wide variety of
nonequilibrium systems that exhibit propagation of well-developed
patterns and fronts into initially unstable and homogeneous states.
Examples occur in such diverse fields as population dynamics
\Ref\fisherkpp{R. A. Fisher, \journal Ann. Eugenics &7&355(37); A.
Kolmogorov, I. Petrovsky, and N. Piskunov, \journal Bull. Univ. Moskou.
Ser. Internat. Sec. A &1&1(37).}  and pulse propagation\Ref\scott{A. C.
Scott, {\sl Neurophysics} (Wiley, New York, 1977); \journal Rev. Mod.
Phys. &47&487(75).}  in nerves in biophysical systems, Taylor-Couette
flows\refmark{\ahlers} and Rayleigh-Benard convection\refmark{
\fineberg} in hydrodynamic systems, crystal
growth\rlap,\refmark{\ben,}\Ref\langer{J. S. Langer and H.
Muller-Krumbhaar, \journal Acta Metall. &26&1681(78); \journal {\sl
ibid.} &26&1689(78); \journal {\sl ibid.} &26&1697(78); J. S. Langer,
\journal Rev. Mod. Phys. &52&1(80).}  models of solidification and
aggregation\rlap,\Ref\nauenberg{M. Nauenberg, \journal Phys. Rev. B
&28&449(83); M. Nauenberg, R. Richter, and L. M. Sander, \journal {\sl
ibid.}  &28&1650(83).}  as well as travelling waves in
reaction-diffusion systems\rlap.\Ref\fife{P. Fife, {\sl Mathematical
Aspects of Reacting and Diffusing systems}, Vol. 28 of {\sl Lecture
Notes in Biomathematics}, edited by S. Levin (Springer, New York, 1979);
J.  Smoller, {\sl Shock Waves and Reaction-Diffusion Equations}
(Springer, New York, 1983).}
In the present paper, we consider only those systems for
which the pattern emerging behind the propagating fronts is homogeneous,
or can be described by an envelope curve, and concentrate on predicting the
propagation velocities of uniformly
translating fronts.

There already exist several proposed
criteria\refmark{\dee,\benjocab,}\Ref\brush{K. V. Brushlinsky and Ya.
M.  Kazhdan, \journal Russ. Math. Surveys &18&1(63).}  for the dynamical
velocity selection mechanism: a minimum speed rule, marginal stability
and structural stability.  Amongst them, the minimum speed principle has
in certain cases a rigorous basis in the Aronson-Weinberger
theorem\rlap.\refmark{\aronson}\ According to the marginal stability
hypothesis\rlap,\refmark{\dee, \benjocab,\shraiman,\saarloos,\van} for
most sufficiently localized initial conditions, the propagation velocity
of well-developed fronts generically approaches the marginal stability
point which apparently coincides with the minimal velocity out of a
family of stable propagating fronts.  The marginal-stability point can
be determined explicitly from the linearized leading edge approximation,
in which only the linearized equation of motion is studied near the
front. In the literature, this is sometimes referred to as the {\it
linear-marginal-stability} case or the {\it pulled} case.  However, as
emphasized by Ben-Jacob \etal\refmark{\benjocab} and later by van
Saarloos\rlap,\refmark{\saarloos,\van} on the basis of rigorous work by
Aronson and Weinberger\refmark{\aronson} for a class of simple
equations, there exist
cases in which the
linear-marginal-stability selection fails.  This is often referred to as
the {\it pushed} case or {\it nonlinear-marginal-stability} case.  Thus
the marginal stability argument apparently provides a practical method
to calculate the selected velocity analytically, but cannot tell when
the method is reliable, since there is no general method known to
distinguish {\it a priori\/} between pushed and pulled cases.  This is
an important limitation, because a given equation may make a transition
from pulled to pushed cases as a parameter in this equation is varied.

The most recent proposal is the structural stability
hypothesis\rlap,\REFS\stri{G.\ C.\ Paquette and Y.\ Oono, University of
Illinois preprint.}\REFSCON\strii{G.\ C.\ Paquette, L.-Y.\ Chen,
N.\ D.\ Goldenfeld, and Y.\ Oono, {\sl Phys. Rev. Lett.} (in
press).}\REFSCON\striii{L.-Y.\ Chen, N.\ D.\ Goldenfeld, Y.\ Oono, and
G.\ C.\ Paquette, {\sl Physica A} (in press).}\refsend in which the
observable fronts are supposed to be stable against small changes in the
governing partial differential equation (PDE) itself.
When the Aronson-Weinberger theorem is applicable,
this hypothesis has the status of a theorem.  The hypothesis applies (so
far) without counterexamples to both the pushed and pulled cases, but
until now it has not been able to yield an analytical means to obtain
the selected velocity.  Instead, a numerical method, based on the structural
stability hypothesis, has been used.

The principal purpose of this paper is to use the perturbative
renormalization group (RG) theory to analyse the stability of
uniformly translating fronts, and to show that this method can be used to
calculate the change in front propagation velocity when the governing
equation of a structurally stable front is perturbed.  In addition, the
RG allows us to predict the uniquely selected velocity by combining the
structural stability principle with it.  As a by-product of our study of
fronts, we have investigated a variational principle due to
Hadeler and Rothe\Ref\hadeler{K. P. Hadeler and F. Rothe, \journal
J. Math.  Biol. &2&251(75).} for the phase
trajectories of steady-state solutions of propagating fronts.
We will see that the variational principle
can give useful upperbounds on propagation speeds in both pulled and
pushed cases, and is able to estimate the transition point between those
regimes.  This work, presented in section 6, is due solely to L.-Y.
Chen.

The applicability of RG to travelling fronts is predicated upon its
recent successful application to the study of long-time global behavior
of physical systems described by nonlinear PDEs: porous medium
equations\rlap,\REFS\goldenfeld{N. Goldenfeld, O. Martin and Y. Oono,
\journal J. Sci.  Comp.&4&355(89).}\REFSCON\oono{N. Goldenfeld, O.
Martin, Y. Oono and F. Liu, \journal Phys. Rev. Lett.
&64&1361(90).}\REFSCON\nato{N. Goldenfeld, O. Martin and Y.  Oono, {\sl
Proceedings of the NATO Advanced Research Workshop on Asymptotics Beyond
All Orders}, S. Tanveer (ed.) (Plenum Press,
1992).}\REFSCON\cgo{L.-Y.\ Chen, N.\ D.\ Goldenfeld, and Y.\ Oono,
\journal Phys. Rev. A &44&6544(91).}\refsend convection-diffusion
transport with irreversible sorption\rlap,\Ref\entov{V.\ M.\  Entov,
I.\  S.\ Ginzburg and E.\ V.\ Theodorovich, \journal J. Appl. Maths. Mechs.
&56&59(92).} the propagation of a turbulent
burst\rlap,\REFS\lin{L.-Y.\ Chen and N.\ D.\ Goldenfeld, \journal
Phys.\  Rev.\ A &45&5572(92).}\REFSCON\pitman{G.\ I.\ Barenblatt, in
{\sl Nonlinear Dynamics and Turbulence}, G.\ I.\  Barenblatt, G.\ Iooss,
and D.\  D.\ Joseph (eds.)(Pitman 1983).}\refsend and linear continuum
mechanics\rlap.\Ref\ndgii{N.\ Goldenfeld and Y.\  Oono, \journal Physica
A &177&213(91).}    In these problems, the global solutions approach the
similarity form $u(x,t) = t^{-\alpha}f(xt^{-\beta})$,
where $x$ is the spatial coordinate, $t$ the time, and $\alpha$ and
$\beta$ are constants.  In many cases -- the so-called intermediate
asymptotics of the second kind\Ref\bar{G.\ I.\ Barenblatt, {\em
Similarity, Self-Similarity, and Intermediate Asymptotics} (Consultants
Bureau, New York, 1979)} -- the exponents $\alpha$ and $\beta$ cannot be
determined by simple dimensional analysis.  In fact, these exponents are
analogues of the anomalous dimensions in field theory, and can be
evaluated by using renormalized perturbation
theory\rlap.\refmark{\goldenfeld-\nato} A mathematically rigorous
formulation has been given by Bricmont, Kupiainen and
Lin\rlap;\Ref\bricmonti{J.\ Bricmont, A.\ Kupiainen and G.\  Lin, {\sl
Renormalization Group and Asymptotics of Solutions of Nonlinear
Parabolic Equations,} Commun. Pure Appl. Math., to appear.}  these
methods have also been used to study front propagation in the
Ginzburg-Landau equation\rlap.\REFS\bricmontii{J.\ Bricmont and A.\
Kupiainen, \journal Commun. Math. Phys.
&150&193(92).}\REFSCON\bricmontiii{J.\ Bricmont and A.\ Kupiainen, {\sl
Stability of Moving Fronts in the Ginzburg-Landau Equation,}
unpublished.}\refsend  A detailed and pedagogical discussion of the
application of RG to the asymptotics of partial differential equations,
and its physical interpretation is given in the book by
Goldenfeld\rlap.\Ref\nigbook{N.D.  Goldenfeld, {\sl Lectures on Phase
Transitions and the Renormalization Group}, (Addison-Wesley, Reading,
Mass., 1992), chapter 10.}  We emphasise only that RG can be used to
solve the above categories of problems; standard techniques, such as
multiple-scale analysis may also be used (perhaps with more difficulty),
and in fact there is a general relationship between multiple-scale
analysis and the RG, whose elements are discussed
elsewhere\rlap.\refmark{\striii}

A steady propagating wave solution has the form $\psi\left(x-ct\right)$,
where $c$ is the propagation speed. If we introduce $X$ and $T$ as
$x=\ln X$, $t=\ln T$, then
$\psi\left(x,t\right)=\Psi\left(XT^{-c}\right)$ for some $\Psi$.  That
is, the solutions can be expressed in the form of similarity solutions.
Because $X$ and $T$ must already be dimensionless, the velocity $c$ can
be viewed as an anomalous dimension, analogous to the exponents
$\alpha$, $\beta$ above in the intermediate asymptotics of the second
kind.  Thus it is a natural guess that there is an RG method to compute
the velocity.  However, there is an important difference between the
similarity solution problems and the propagation problem.  In the
former, there is a unique asymptotic exponent, whereas in the latter the
propagation speed is not unique.  Thus, for the selection problem RG is
not sufficient.  We will see that it must be combined with some new
physics --- the structural stability hypothesis.

The outline of this paper is as follows.  In Section 2 a formal
renormalized perturbation theory for the front propagation is
formulated.  In Section 3, technical aspects of the renormalization
method are discussed.  The central issue is the nature of the eigenvalue
spectrum of the linearised PDE; this is resolved using considerations of
structural stability.  In Section 4, we discuss the relationship between
solvability conditions and renormalizability.  Section 5 contains a
number of applications of the formalism, in which we compute the
propagation velocity using perturbative RG.  Section 6 presents results
using the variational principle and a simple trial function.  In section
7, we use the variational and RG methods to predict the transition
between the pushed and pulled cases.  We summarise and conclude in
section 8.

{\bf \chapter{Formal Theory of Perturbative Renormalization}}

In this paper we mostly concentrate on the so-called
Fisher-KPP equation\rlap:\refmark{\fisherkpp}
$$
\frac{\partial \psi}{\partial t} = \frac{\partial^2\psi}{\partial x^2}
+ F ( \psi ),    \eqn\fisher
$$
where $F$ is a continuous function with $F(0) = F(1) = 0$.  If $F$
satisfies the condition $ F(\psi) > 0$ for all $\psi \in (0,1)$, then
there exists a stable traveling-wave solution interpolating between
$\psi = 1$ and $\psi = 0$ with propagation speed $c$ for each value of
$c$ greater than or equal to some minimum value $c^*$, where $c^*  \ge
\hat{c} \equiv2 \sqrt{F'(0)}$ (if $F$ is differentiable at the origin).
As suggested by the analogy between similarity solutions and travelling
waves, there should be an RG method to compute the propagation speed.
Here we show that when a front is structurally stable, we can devise a
renormalized perturbation method to compute the change in the
propagation speed due to a small perturbation in \eg\ the `reaction
term' $F$.
\bigskip
{\bf \cl{A. Perturbation Theory}}

We consider formally a system described by the following abstract
nonlinear (usually parabolic) equation:
$$
\frac{\pd \psi}{\pd t}=N\{\psi \}.\eqn\rgone
$$
Let $\psi_{0}(x - c_{0}t + x_{0})$ be a stable traveling front solution
of \rgone\ with speed $c_{0}$, where $x_{0}$ is a free parameter
indicating the translational symmetry of the problem explicitly; there
is a one-parameter family of traveling wave solutions with a given
speed.  Let us add a small structural perturbation $\delta N$ to
\rgone; for example in
\fisher\ we replace $F$ with $F + \delta F$.
We assume that the operator norm of $\delta N$ on an appropriate domain
is less than some small positive number $\epsilon$; for example for
\fisher, the $C^{0}$-norm of $\delta F$ is less than $\epsilon$: that
is, $||\delta F|| < \epsilon$, where the norm $|| \delta F || \equiv
\sup_{u} | \delta F(u) |$.  Assume that in response the front solution
is modified to $\psi_{0} + \delta \psi$, where
$||\delta\psi||$ is of order $\epsilon$.  Linearizing \rgone\ to order
$\epsilon$ in the moving frame
with velocity $c_{0}$, we find the following equation governing the
first order correction
$$
\frac{\pd \delta \psi}{\pd t}=\left(c_0\frac{\pd }{\pd \xi}+DN\right)
\delta \psi+\delta N\{\psi_0\},\eqn\rgthree
$$
where $DN $ is the Fr\'{e}chet derivative of $N$ at $\psi = \psi_{0}$,
$\xi\equiv x -c_0 t + x_0$ and the initial condition satisfies $\delta
\psi(\xi,t_0)=0$.

To make our mathematical formalism as simple as possible, first, without
loss of generality, we will consider the case for which the highest
differential operator is the second order; however, the conclusions to be
drawn below apply to other high-order nonlinear operators too.

The formal solution to equation \rgthree\ reads
$$
\delta \psi(\xi,t)=\int_{t_0}^{t}dt'\int_{-\infty}^{+\infty}d\xi'
G(\xi,t;\xi',t')   [\delta N\{u_0\}](\xi'),\eqn\rgfour
$$
where $G$ is the Green function,
which satisfies
$$
\frac{\pd G}{\pd t}-\left(c_0\frac{\pd }{\pd \xi}+DN\right)G
=\delta(t-t')\delta(\xi-\xi')\eqn\rgfive
$$
with $G \rightarrow 0$ in $|\xi - \xi'| \rightarrow \infty$.
The Green function may formally be written as
$$
G(\xi,t;\xi',t')=\sqrt{\rho(\xi')\rho(\xi)}
\sum_{n=0}^{\infty}\frac{1}{C_n^2}e^{-\omega_n
(t-t')}u_n(\xi)u^*_n(\xi')\Theta(t-t'),\eqn\rgsix
$$
where we use the summation symbol to represent both summation and
integration; $\rho(\xi)$ is an appropriate weight function,
$\{C_n^2\}$ are appropriate normalization constants,
and $\Theta$ is the Heaviside function.  The summation is
over the spectrum $\{ \omega_n \}$, and $\{ u_n \}$ are corresponding
appropriately normalized (generalized) eigenfunctions obeying
$$\left(c_0\frac{d}{d \xi}+DN\right)u_n=-\omega_n u_n.\eqn\rgseven$$
We do not need to solve
completely this seemingly formidable equation.  Differentiating equation
\rgone\ with respect to $\xi$, we have
$$
\left(c_0\frac{d }{d\xi}+DN\right)\left(\frac{d\psi_0}{d\xi}\right)=0,
\eqn\rgeight$$
showing that, due to the translational invariance of the original
equation, $u_0 \propto d\psi_0/d\xi$ satisfies \rgseven\.  For the
fronts in which we are interested, this function is well localized
(square integrable), so that $0$ is an eigenvalue.   We assume that the
operator on the left hand side of \rgseven\ is dissipative, so that $0$
is the upper bound of its spectrum.

Thus, only the zeroth eigenfunction $\psi_0$ contributes to
the secular term which is proportional to $t-t_0$, and
the perturbed solution can be written as
$$
\psi(\xi,t)=\psi_0(\xi)- \delta c (t-t_0) \frac{d\psi_0}{d\xi}
+\left(\delta \psi\right)_r + O(\epsilon^{2}).
\eqn\bare
$$
Here $\delta c$ is given by
$$
\delta c=-\frac{\int_{-\infty}^{+\infty}d\xi\rho(\xi)u_0(\xi) [\delta
N\{u_0\}](\xi')}{\int_{-\infty}^{+\infty} d\xi\rho(\xi)u_0^{2}(\xi)}
,\eqn\rgtwelve
$$
and $(\delta \psi)_r$ represents the non-secular terms, which are
unimportant for the purpose of determining the modification of the
velocity, leading only to finite $O(\epsilon)$ corrections to
the profile of the front.

One may immediately guess that this expresssion for $\delta c$
is the change in the front speed, but there are two problems
with this identification.  First,
the naive perturbation theory is
not controlled due to the secularity.  A renormalization procedure,
given below, will be used to remove the secularity.  Second,
both
the numerator and the denominator of the expression for $\delta c$ may
be divergent.  This potential difficulty
may be treated using the considerations of structural stability given
in section 3.
\bigskip
{\bf \cl{B. Renormalization and Renormalization Group}}

For a certain class of sufficiently localized initial conditions, all
the transient solutions to the unperturbed PDE of the form
$u(x,t)=u(x-x_c(t,c_0,x_0),t)$, asymptotically converge to the same
uniformly translating propagating front, $u(x,t)=u(x-c_0 t+x_0)$ as
$t\rightarrow \infty$.  After the perturbation has been switched on,
$x_0$ is no longer a constant of motion of the perturbed system, and is
therefore not observable at long times.  From a measurement of $x_0$ at
late times, the initial value of $x_0$ cannot be deduced; hence, it must
be renormalized from the RG point of view\rlap.\refmark{\nigbook}\ As we
saw in the introduction, if we make transformations $x=\log X, t=\log T,
x_0=\log A_0$, and $u(x,t)=U(X,T)$, the travelling wave solution can be
expressed in the form of a similarity solution $U(X,T)=U\left(\log A_0
X/ T^{c_0}\right)=U_1\left(A_0 X/T^{c_0}\right)=U_2\left(T/(A_0
X)^{1/c_0}\right)$, for appropriate functions $U$, $U_1$ and $U_2$.  It
now becomes clear that the role of the constant of motion $x_0$ is the
same as that played by the initial total amount of mass or energy $Q_0$
in the nonlinear diffusion problems studied
previously\rlap.\refmark{\bar-\pitman}\ Although it is possible to
renormalize travelling wave problems by regarding them as similarity
solutions\rlap,\Ref\usunpub{L.-Y. Chen, Nigel Goldenfeld, Y. Oono,
unpublished.}  here we will perform a new type of RG analysis directly
on the travelling wave solution.

The divergence of the secular term $(\delta u)_s\propto (t-t_0)$ as $t
\rightarrow \infty$, can be removed order by order in $\epsilon$
by regarding $t_0$ as a regularization
parameter and
introducing an additive renormalization constant $Z=Z(t_0, \mu, \epsilon)$
with $\mu$ an arbitrary time.
Note that
the limit $t-t_0 \rightarrow \infty$ can be achieved in two equivalent ways:
either keeping $t_0$ fixed and letting $t \rightarrow \infty$, or keeping
$t$ fixed and letting $t_0 \rightarrow -\infty$. We will use the latter method
in the following text, which corresponds to the conventional treatment
of logarithmic divergences in the zero
cutoff limit of the bare perturbation series for (\eg)
$\varphi^{4}$ field theory. $t$ corresponds to $\ln T$ in the analogy
discussed in the preceeding section, so that $t-t_{0}$ corresponds to
$\ln(T/T_{0})$, where $\ln T_{0}$ corresponds to $t_{0}$.  In the
standard  Gell-Mann--Low
RG\rlap,\Ref\gell{M.\ Gell-Mann and F.\ E.\  Low, \journal Phys. Rev.
&95&1300(54).} we introduce an arbitrary length scale (or rather time
scale in this context) $L$ and split $T/T_{0}$ into $T/L\cdot L/T_{0}$;
the term $\ln (T_{0}/L)$ is absorbed into the (multiplicative)
renormalization constant often denoted by $Z$.  The corresponding
procedure here is to split $t-t_{0}$ into $t-\mu - (t_{0}-\mu)$, where $\mu$
corresponds to $\ln L$ in the standard approach.  Since multiplications
in the standard multiplicative renormalization procedure correspond to
addition in our case, we renormalize $x_0(t_0)$ by
$x_{0}^{R}+ Z(t_0, \mu, \epsilon)$, and Taylor expand
$Z=\sum_{n=1}^{\infty}
a_n(t_0, \mu)\epsilon^{n}$. The coefficients $a_n (n \ge 1)$ are
determined order by order in $\epsilon$ in such a way that all the secular
divergences in $\psi(x,t)$ are cancelled out.
In this way, the renormalized solution $\psi(x,t)$ remains finite even in
the limit $t_0 \rightarrow -\infty$.
Hence, to $O(\epsilon)$, the
solution $\psi(x,t)$ can be rewritten as
$$\eqalign{
\psi(x,t)
= & \psi_0\left(\xi_0\right)-\delta c (t-t_0) \frac{d\psi_0}
{d\xi_0}+O(\epsilon),\qquad  \xi_0=x-c_0 t+x_0 \cr
= & \psi_0\left(\xi+\epsilon a_1 \right)-\delta c (t-t_0) \frac{d\psi_0}
{d\xi}+O(\epsilon),\qquad  \xi=x-c_0 t+x_0^R \cr
= & \psi_0\left(\xi\right)+\epsilon a_1  \frac{d\psi_0}{d\xi}-\delta c
(t-t_0) \frac{d\psi_0}{d\xi}+O(\epsilon),\cr}
\eqn\rgthirteen$$
where $O(\epsilon)$ refers to all finite terms
of order $\epsilon$, regular in the limit $t_0 \rightarrow -\infty$.
By choosing
$a_1 \epsilon =\delta c (\mu-t_0)$, the secular divergence is removed.
We obtain the renormalized solution
$$\psi(x,t)=\psi_0\left(\xi\right)- \delta c (t-\mu)\frac{d\psi_0}
{d\xi}+ O(\epsilon), \qquad \xi=x-c_0 t +x_0^R (\mu).\eqn\ranone$$
It is impossible that the actual solution $\psi(x,t)$ can depend on the
arbitrary
time scale $\mu$, because (as with $L$ in the standard procedure) $\mu$ is not
present in the original problem.
This is expressed by the renormalization group equation
$$\frac{\pd \psi}{\pd \mu}{\Big |}_{x, c_0, t, \epsilon}=0.\eqn\rantwo$$
Hence, to
order $\epsilon$ the RG equation yields, after equating $\mu$ with $t$,
$$
\frac{\partial \psi}{\partial t} + \delta c \frac{\partial \psi}{\partial \xi}
= 0.   \eqn\rgeq$$
This has the form of an amplitude equation, an observation
discussed further in section 4, and in ref. \striii.
Thus the speed of the renormalized wave is indeed $c=c_{0} + \delta c$, and
the leading long-time asymptotic behavior, to
$O(\epsilon)$, is
$$\psi(x,t)\sim \psi_0(x-ct+x'_0)+O(\epsilon),\eqn\rgfourteen$$
where $x'_0$ is the new constant of motion for the perturbed system.
\bigskip
{\bf \cl{C. Application to Fisher's Equation}}

Now we consider the application of the above formalism to Fisher's
equation.  An important technical aspect of this discussion is deferred
to section 3.  We consider only a perturbation $\delta F$
to the nonlinear reaction
term $F$ of \fisher\, so that \rgfour\ reads
$$
\delta\psi(\xi_{0},t)= e^{-c_{0}\xi_0 /2}\int_{t_0}^{t}dt'
\int_{-\infty}^{+\infty}d\xi' G(\xi_{0},t;\xi',t') e^{c_{0}\xi'/2}
[\delta F\{\psi_0\}](\xi').\eqn\rgfour
$$
Here, $G$ is the Green function satisfying
$$
\frac{\partial G}{\partial t}- {\cal L}G
=\delta(t-t')\delta(\xi-\xi')
$$
with $G \rightarrow 0$ in $|\xi - \xi'| \rightarrow \infty$, and
$$
{\cal L} \equiv \frac{\partial^{2}}{\partial \xi^{2}} +
F'(\psi_{0}(\xi)) - \frac{c_{0}^{2}}{4}.
$$
Formally, $G$ reads
$$
G(\xi,t;\xi',t') =  u_{0}(\xi)u_{0}^{*}(\xi') + \sum
e^{-\lambda_{n}(t-t')} u_{n}(\xi) u_{n}^{*}(\xi'),
$$
where ${\cal L} u_{0} = 0$, and ${\cal L} u_{n} = \lambda_{n} u_{n}$.
The summation symbol, which may imply appropriate integration, is over
the spectrum other than the point spectrum $\{ 0 \}$.  Since the system
is translationally symmetric, $u_{0} \propto
e^{c_{0}\xi/2}\psi_{0}'(\xi)$.  Due to the known stability of the
propagating wave front, the operator ${\cal L}$ is dissipative, so that
zero is the least upper bound of its spectrum.  Hence, only $u_{0}$
contributes to the secular term in $\delta \psi$.  A similar argument as
in the general case gives the explicit formula for
\rgtwelve\:
$$
\delta c = - \frac{
\int_{-\infty}^{+\infty}d\xi e^{c_{0}\xi}\psi'_{0}(\xi) \delta
F\{\psi_{0}\}(\xi)  }{ \int_{-\infty}^{+\infty} d\xi e^{c_{0}\xi}
\psi_{0}^{'2}(\xi) }. \eqn\dc
$$

{\bf \chapter{Justification of Perturbative RG and Structural Stability}}

Before proceeding further, we must now address the issue that the
formula for $\delta c$ may not be finite, for example.  The formal
perturbation approach may fail if $0$ is not an isolated eigenvalue of
the linearized operator.  In this section, we will discuss when the
restricted formula \dc\ is meaningful.  To this end, we need results
from our structural stability analysis, and so we begin with a brief summary
of this topic\rlap.\refmark{\stri-\striii}

Equation \fisher\ can be classified into two cases: ambiguous and
unambiguous.  We say that the equation is {\em unambiguous} if it allows a
unique propagating speed for its traveling wave solutions (here, we
consider the waves traveling in the positive direction only), and is
ambiguous otherwise.  A necessary and sufficient condition for
\fisher\ to be unambiguous is that $F$ does not have any isolated
minimum at the origin.  Notice that by a certain indefinitely small
$C^{0}$-perturbation $\delta F$ of $F$, we can convert an ambiguous
equation into an unambiguous one.  Hence, at most one
propagating speed can be stable against this modification of the
system.  Because we are perturbing $F$, that is, the equation itself, we
call such a perturbation a structural perturbation.  Thus we may say
that there is at most one structurally stable propagating speed for
\fisher\.  It has been proved that the slowest propagating speed $c^*$
which allows stable (in the usual sense) propagating waves is a
continuous functional of $F$ so long as $Q(F) \equiv \sup_{\psi \in
\left(0,1 \right]} F(\psi)/\psi$ is a continuous functional of $F$.
This is satisfied, if $\delta F$ is $C^{0}$-small and   $\sup_{\psi \in
\left(0,1 \right]} \delta F(\psi)/\psi$ is smaller than some positive
number which converges to zero as $||\delta F||$ goes to zero.  We call
such perturbation a physically small ($p$-small) perturbation.  Thus we
may say that the wave with speed $c^*$ is structural stable against
$p$-small structural perturbations.

We have conjectured that only structurally stable solutions of a model
equation correspond to the physically observable phenomena.  This
structural stability hypothesis is closely related to, but is not
identical with the idea originally proposed by Andronov and
Pontrjagin\Ref\andro{A.\ Andronov and L.\ Pontrjagin, \journal
Dokl.\ Akad.\ Nauk.\ SSSR &14&247(1937)} for dynamical systems.  The
conjecture implies for \fisher\ that $c^*$ is the selected propagation
speed.  It seems to be widely believed that $c^*$ is indeed the unique
observable speed for \fisher\, but a proof does not exist for general
$F$\rlap.\Ref\aronwein{D.\ G.\ Aronson and H.\ F.\ Weinberger, in {\it
Partial Differential Equations and Related Topics}, edited by
J.\ A.\ Goldstein (Springer, Heidelberg, 1975); H.\ F.\ Weinberger,
\journal SIAM J. Math. Anal. &13&3(1982).}  However, our numerical
studies have so far failed to uncover counterexamples.  The structural
stability hypothesis is usually redundant, but has predictive power in
cases where one's model of a physical phenomenon inadvertently includes
unphysical features.  This is the case for the over-idealised models
(such as \fisher) of propagation phenomena considered here.

Prompted by our structural stability analysis, we have used
the following numerical method to estimate the selected velocity.  Apply the
shooting method to the ordinary differential equation (ODE) governing
the propagating wave front shape $\psi$:
$$
\frac{d^{2} \psi}{d \xi^{2}} = - c\frac{d \psi}{d \xi} +
F(\psi), \eqn\ode
$$
where $c$ is the propagating speed, and $\xi$ corresponds to $ x - ct$,
after modifying $F$ by adding a $p$-small perturbation to convert the
equation into the unambiguous class.  We know that the unique speed is almost
identical to the true selected velocity, since $c^*$ is continuously
dependent on $F$.  Therefore, there is only one $c$ which allows the
shooting method to give a solution.  This $c$ is the selected speed.
However, in practice, the method may not be very accurate, because the true
unique solution and other solutions which do not reach $\psi = 0$
for large $\xi$ may not be easy to distinguish when the perturbations are
small.

There are several important consequences of our structural stability
study.  First, notice that even if $\delta F$ is $p$-small, $-\delta F$
need not be.  Furthermore, it is easy to construct an indefinitely
$C^{0}$-small $\delta F$ which increases $F'(0)$ indefinitely, but
$-\delta F$ is $p$-small.  An example of such $\delta F$ is a very tiny
but very sharp spike well localized near the origin.  This implies that
for this $\delta F$, $\delta c$ must be indefinitely large, while
$\delta c$ for $-\delta F$ is infinitesimal.  Now \dc\ is a linear
functional of $\delta F$, so that the sign change of $\delta F$ cannot
cause such a drastic change.   This clearly demonstrates that the
formula \dc\ is not meaningful, if either $\delta F$ or $-\delta F$ is
not $p$-small.  More explicitly, $\sup_{u}|\delta F(u)/u|$ must vanish
as $||\delta F||$ goes to zero.
If $\delta F$ is differentiable
at the origin, the condition is satisfied.

The perturbation approach outlined above is legitimate only when 0 is an
isolated point spectrum of the operator ${\cal L}$.  Unfortunately, this
is not the case for many reaction diffusion equations.  For \fisher\ the
essential spectrum has the range $(-\infty, \eta)$, where $\eta =
\max\{F'(1), F'(0)\} - c_{0}^{2}/4$, as is easily seen from Rota's
theorem (a generalization of Weyl's
theorem)\rlap.\Ref\rota{G.\ C.\ Rota, \journal Commun.\ Pure
Appl.\ Math. &11&23(1968).}  Hence, if $c_{0} =2 \sqrt{F'(0)}$, which is
the important case for the pulled equations, then the eigenvalue 0 is
not isolated.  Nevertheless, we can justify the formula or rather its
augmented version as follows.

Notice that we can always find a sequence $\{\delta f_{k}\}$ of
piecewise differentiable perturbations  such that
$\delta f_k$ converges
to zero in $C^{0}$-norm, but $\delta f'_{k}(0)$ is always $-1$.  This is
a sequence of $p$-small perturbations, so that $c^*(F+\delta f_{k})$
converges to $c^*(F)$.  However, the essential spectrum of the operators
${\cal L}$ is bounded from above by $-1$, so for these perturbed
systems, 0 is always isolated.  Now consider the perturbed system with
$\delta F$.  Instead of the system with $F + \delta F$, we again
consider a sequence of systems $F + \delta F + \delta f_{k}$.  For this
sequence, $c^*$ converges to $c^*(F + \delta F)$.  Thus we may study the
perturbation of the system with $\delta F + \delta f_{k}$ instead of the
original system with $\delta F$ to compute the change in velocity.  Thus
we may conclude that the ordinary perturbation theory can be used as
done formally in the preceding section.

For the pulled case, $c^* = 2\sqrt{F'(0)}$, so that both the numerator
and the denominator of \dc\ diverge.  This difficulty is removed if we
consider the sequence mentioned above instead of the original system.
In this case all the members of the sequence are unambiguous equations
(pushed cases), so the decay rate of the eigenfunction belonging to 0
is much faster than $e^{c^{*}\xi}$ near the tip.  That is, we have a
natural regularizing factor for the members of the sequence.  Taking
the limit is thus equivalent to
computing the limit with the aid of
l'H\^opital's rule:
$$
\delta c = - \lim_{\ell \rightarrow \infty}\frac{
\int_{-\ell}^{+\ell}d\xi e^{c_{0}\xi}\psi'_{0}(\xi) [\delta
F\{\psi_{0}\}](\xi)  }{ \int_{-\ell}^{+\ell} d\xi e^{c_{0}\xi}
\psi_{0}^{'2}(\xi) }. \eqn\dcl
$$

Thus with the aid of the structural stability consideration, we can
give the correct form \dcl\ for the change of velocity and a sufficient
condition for its validity: $\pm \delta F$ must be $p$-small and
the right and the left derivatives at the origin must exist.

{\bf\chapter{Amplitude Equations, Solvability and RG}}

The general formula \rgtwelve\ for the modification of velocity
due to the external perturbations can also be constructed from a
solvability condition, analogous to that occuring in
the dynamics of defects and
dislocations in nonequilibrium patterns\rlap.\REFS\siggia{E. D.
Siggia and A. Zippelius, \journal Phys. Rev. A
&24&1036(81).}\REFSCON\pomeau{Y. Pomeau,  S. Zaleski, and P. Manneville,
\journal Phys. Rev. A  &27&2710(83).}\REFSCON\cross{G. Tesauro and M. C.
Cross, \journal Phys. Rev. A &34&1363(86).}  \REFSCON\bod{E.
Bodenschatz, W. Pesch, and L. Kramer, \journal Physica D
&32&135(88).}\refsend
We start with the perturbed equation
$$
\frac{\pd \psi}{\pd t}=N\{\psi\} + \delta N\{\psi\},\eqn\rgfirst
$$
which is assumed to have a steady-state solution of the propagating
front $\psi(x,t)=\psi(\xi)$, with $\xi=x-c_0 t+x_0$.  We further assume
that the operator norm of $\delta N$ is of order $\epsilon$.  Suppose
that the perturbed velocity and solution can be written as $c=c_0+
\delta c$ and $\psi=\psi_0+ \delta \psi$, where $c_0$ and $\psi_0$
correspond to the unperturbed velocity and solution.   Then to order
$\epsilon$ we get
$$
-\left(DN + c_0\frac{d }{d\xi}\right) \delta \psi =\delta c \frac
{d\psi_0}{d\xi}+\delta N\{\psi_0\}.\eqn\rgthird
$$
Because the linear operator on the left hand side of equation
\rgthird\ has a zero eigenvalue corresponding to an eigenfunction
$u_0\propto d\psi_0/d\xi$, the condition for the existence of a
nontrivial solution to \rgthird\ is that the right-hand side should be
orthogonal to the null space of the operator: $\tilde {\psi_0}\equiv
\rho(\xi) \psi_0=\rho(\xi) du_0/d\xi$, where we have included the
appropriate weight function $\rho$:  $$\delta c\left(\rho
\psi_0,\psi_0\right)+\left(\rho \psi_0,\delta N\{\psi_0\} \right)=0
.\eqn\rgfourth$$ We immediately recover the formula \rgtwelve\ for
$\delta c$.

In particular, let us now consider a case in which the change in the speed
$\delta c$
results from the change of some parameter $\gamma$ in the nonlinear operator
$N(\psi,\gamma)$. If we assume that the traveling wave solution exists for
some $\gamma$, and that we can write $\delta N=({d N}/{d \gamma})
\delta \gamma$, and further assume that
the order of the differentiation and the integration can be exchanged,
then from \rgtwelve\,
we obtain the formally exact result
$$
\frac{d c}{d \gamma} = - \frac{
\int_{-\infty}^{+\infty}d\xi e^{c\xi}\psi'_{0}(\xi)
d N/d \gamma }{ \int_{-\infty}^{+\infty} d\xi e^{c\xi}
\psi_{0}^{'2}(\xi) }. \eqn\sol
$$
Unfortunately, we have not found a way to use this formula in cases
where exact results are not already available.

The reader may well wonder why we make a fuss about the RG in
the preceding section, given that the formula for $\delta c$ is obtained
trivially from the solvability condition of \rgthird\. Our point is
to demonstrate that the solvability conditions (order by order) and the
perturbative renormalization procedure are equivalent. The use of
solvability conditions is reminiscent of pattern selection in
dendritic crystal growth phenomena\rlap,\refmark{\ben} the derivation of
amplitude equations, such as the time-dependent Ginzburg-Landau equation, or
more generally the equations of motion on the slow or center manifold
(in the loose sense of the word)\rlap.\Ref\crawford{A recent review
is given by J.D. Crawford, \journal Rev. Mod. Phys. &63&991(91).}
Indeed, \rgeq\ is the
equation of motion governing the slow motion relative to the original
unperturbed front.  It is actually a general feature that amplitude
equations and slow motion equations are RG equations. These general
features of renormalization group approach are discussed
more thoroughly elsewhere\rlap.\refmark{\striii, \usunpub}

{\bf \chapter{Examples}}

We now apply the perturbation approach to a variety of front propagation
problems.
\bigskip
{\bf \cl{A. Generalized Fisher's Equation.}}

Consider the model:
$$
\frac{\pd u}{\pd t}=\frac{\pd^2 u}{\pd x^2}+u(u-\mu)(1-u),\quad 0\leq
\mu\leq 1/2.\eqn\rgfifteen
$$
Here, we identify $F = u^{2}(1-u)$ and $\delta F = - \mu u (1-u)$ and
regard $\mu$ as a small perturbation parameter.  The equation is
unambiguous for all $\mu \ge -1/2$.  In particular, the unperturbed case
is unambiguous and its unique propagation front shape is described by
$$
u_0=\frac{1}{1+e^{1/\sqrt{2}\xi}},\quad
\xi=x-c_0t+x_0,\eqn\rgsixteen$$
%
%
%
where $c^*\equiv c(\mu=0)=1/\sqrt{2}$ and $x_0$ is an arbitrary constant.  From
\dc\ we obtain $\delta c=-\sqrt{2}\mu$. Thus, the selected velocity of
the perturbed equation, to $O(\mu)$, is $c^*=1/\sqrt{2}-\sqrt{2}\mu$,
which is the same as the exact result from the variational method we will
describe below.
Presumably higher order terms in $\mu$ will spoil this result.
\bigskip
{\bf \cl{B. Van Saarloos' Fourth Order Equation.}}

Van Saarloos considered the following equation\rlap, \refmark{\van}
$$
\frac{\pd \phi}{\pd t}=\frac{\pd^2 \phi}{\pd x^2}-\gamma \frac{\pd^4 \phi}
{\pd x^4}+\frac{\phi}{b}(b+\phi)(1-\phi),\eqn\rgfifty
$$
where $\gamma<1/12$ and $0\leq b\leq 1$. He
numerically solved both this full time-dependent equation for $\psi(x,t)$
and the corresponding ordinary differential equation
for the steady propagating wave state $\phi(\xi)$ in the moving frame with
velocity $c$ for various values of $b$ and $\gamma$\rlap.\refmark{\van}\
For $\gamma=0.08$ and $b=0.1$, he observed that the velocity
$c\simeq 2.715$.

Here we present analytical results obtained by treating the fourth-order
term $-\gamma {\pd^4 \phi}/{\pd x^4}$ as a small perturbation.  The
unperturbed equation has the propagating-front solution\refmark{\van}
$\phi_0(\xi)=(1+e^{\kappa\xi})^{-1}$, $\xi=x-c_0t+x_0$, where
 $\kappa=[c_0-\sqrt{c_0^2-4}]/2$, and $c_0=2$ for $1/2<b<1$;
$\kappa=[c_0+\sqrt{c_0^2-4}]/2$, and $c_0=\sqrt{2b}+1/\sqrt{2b}$, for
$0<b\leq 1/2$. The transition point from
the pulled to
the pushed case is $b=b_c=1/2$. It is worth mentioning that
it is straightforward to obtain these results by applying the
variational method presented in section 6. From the formula \rgtwelve\,
we obtain, for $0<b\leq 1/2$,
$$\delta c=-\gamma \kappa^3\left(1-(2-\frac{c_0}{\kappa})\left(\frac{7}{2}-
\frac{9}{5}(3-\frac{c_0}{\kappa})+\frac{1}{5}(3-\frac{c_0}{\kappa})
(4-\frac{c_0}{\kappa})\right)\right),\eqn\rgfi$$
and $\delta c=-\gamma$ for $1/2<b<1$.

Note that the linear-marginal-stability velocity $c_l$ switches from
$c_l=2$ to $c_l\approx 2-\gamma \approx 2\sqrt{1-\gamma} $ as $b$
crosses the value 1/2.
We find that the $O(\gamma)$ RG prediction agrees very well with the
numerical calculation. For instance, for $\gamma=0.08$, we obtain
$c= 2.696 + O(\gamma^2)$ for $b=0.1$, which is close to the numerical result
$c\approx 2.715$ by van Saarloos\rlap.\refmark{\van}
\bigskip
{\bf \cl{C. Newman's Population Equation.} }

The second example which we consider is a modified porous medium
equation which has been extensively discussed by
Newman\rlap,\Ref\newman{W. I. Newman, \journal J. Theor. Biol.
&85&325(80); \journal PAGEOPH &121&417(83).}  in the contexts of
population genetics and combustion. The equation can be represented as
$$
\frac{\pd u}{\pd t}=\frac{1}{2}\frac{\pd }{\pd x}\left(u^n\frac{\pd u}{\pd x}
\right)+ u(1-u), \quad n\geq 0.\eqn\rgseventeen$$
For $n=0$, it reduces to the Fisher-KPP equation and the propagation velocity
is $c_0=\sqrt{2}$. For $n=1$, it has a unique travelling-wave solution
$$u_0(x,t)=\left(1-e^{\xi}\right)\Theta(-\xi),\quad \xi=x-c_0t+x_0,\quad c_0=
\frac{1}{2},\eqn\rgeigteen$$
where $\Theta$ is the Heavyside function.

Numerical results by Newman \refmark{\newman} suggest that for any
$n>0$, solutions evolve asymptotically into travelling-wave solutions
with a unique velocity $c\approx (n+1)^{-1}$, depending only on the
value of $n$.

To study this problem analytically, we write $n=1+\delta$ and perform a
perturbative RG calculation, regarding $\delta$ as a small parameter.
Expanding equation \rgseventeen\ in $\delta$, we obtain, to $O(\delta)$,
the perturbation $\delta N\{u_0\}=(\delta/2)\,\partial_x\left(u_0 \partial_x
u_0 \log u_0\right)$. The weight function is $\rho(\xi)=e^{\xi}$ and the
ground state is $\psi_0(\xi)=du_0/d\xi=-e^{\xi}, -\infty<\xi\leq 0$.
Using formula \rgtwelve\, we have $\delta c=-(13/48)\, \delta$, and the
perturbed velocity to $O(\delta)$ is $c=1/2-(13/48)\,\delta$.
A consistent formula $c=\left(2+(13/12)\,\delta+O(\delta^2)\right)
^{-1}$ works remarkably well even for $\delta$ as large as one.
Setting
$\delta=1$, \ie\ $n=2$, we obtain $c\approx 0.3243$, which is in
excellent agreement with the numerical result $c\approx 0.32$ obtained
by Newman\rlap.\refmark{\newman}
\bigskip
{\bf \cl{D. Pulled Case Revisited } }

The renormalized perturbation theory result \rgtwelve\ can also be used
to calculate heuristically the selected velocity of the unperturbed
system by imposing the structural stability principle. Within the
perturbation theory, a necessary and sufficient condition that $c^*$ be
the selected speed is that $\delta c(c^*)$
must vanish as $\delta F$ vanishes.  For
concreteness, we consider the
Fisher-KPP equation \fisher\ with $F=\psi
(1-\psi)$. It is found that the change in the velocity $\delta c(c)$ is
zero as $||\delta F|| \rightarrow 0$ for all perturbations $\delta F$,
which are both $p$-small and differentiable at the origin, only for
$c=c^*=2$; for $c>c^*$ there exist such perturbations for which $\delta
c$ does not vanish as $||\delta F|| \rightarrow 0$. A simple example of
the latter is the perturbation $\delta F=\psi
(1-\psi)-(\psi-\Delta)(1-\psi)
\Theta(\psi-\Delta)$, where $\theta$ is
the step function, and we let $\Delta \rightarrow 0^+$.  The RG
calculation shows that $\delta c \sim \sqrt{c^2-4}$ as $\Delta
\rightarrow 0^+$. Obviously, only for $c=c^*=2$, $\delta c$ goes to
zero, but for $c>c^*=2$, $\delta c$ does not vanish. Therefore, only the
wave with $c=c^*=2$ is structurally stable, and $c=2$ is identified as
the selected velocity of the unperturbed system.

{\bf \chapter{Variational Principles}}

In this section, we digress briefly to report results of L.-Y. Chen
using the variational principle of Hadeler and
Rothe\rlap.\refmark{\hadeler}\ With relatively little effort, useful
estimates for front speeds and transition points are obtained.

We consider \fisher\ in the variable $u(x,t)$, where the reaction term
$F$ is continuously differentiable and satisfies the condition
$F(0)=F(u_s)=0$ (where $u=0$ and $u_s$ are two steady-state solutions of
the PDE); other generic conditions will be imposed later.  Hadeler and
Rothe's principle is that\refmark{\hadeler}
$$c_0=\inf_{\rho} \sup_{0\leq u\leq u_s}\{\rho'(u)+\frac{F(u)}{\rho(u)}\}
,\eqn\vpsix$$
where supremum means least upper bound \etc, $'$ denotes differentiation
with respect to the argument, and the function $\rho (u) \equiv
-\pd_x u$ satisfies the conditions
$$\rho(u)>0,\quad 0<u<u_s;\quad \rho(0)=\rho(u_s)=0,\quad \rho'(0)>0,\quad
\rho'(u_s)<0.\eqn\vpfive$$

It was found empirically that if $F$ is of the form
$F(u)=u(u_s^n-u^n)\Phi(u)$, where $\Phi(u)$ is some rational function
which does not have zeros at $u=0$ or $u_s$, then the following choice
very often gives the correct $c^*$:
$$
\rho(u)=u(u_s^n-u^n)\psi(u,\{\kappa\}),\eqn\vptwysix
$$
where $\psi$ is a function positive for $0\leq u\leq u_s$ and
$\{\kappa\}$ is a set of variational parameters.   Often one parameter
$\kappa$, which is the scaling factor of $\rho$, is sufficient.   Even
when the choice fails to give the exact $c^*$, it gives a very good
upper bound of $c^*$.
\bigskip
{\bf \cl{A. Fisher's Population Model.}}
\medskip

We consider the equation $$\frac{\pd u}{\pd t}=\frac{\pd^2 u}{\pd x^2} +
u(1-u)(1+\nu u^n),\quad n\geq 1, \quad -1\leq \nu <
+\infty,\eqn\vpseven$$ where $u=0, 1$ are two steady-state points. When
$n=1$, (2.7) reduces to the original Fisher's population model that
Hadeler and Rothe discussed in Ref.  \hadeler.  They indirectly obtained
correct results for $n=1$ by taking advantage of the known exact
solution of the front profile although they did not choose a proper trial
function. Here, we show that, for $n> 1$, the velocities and values of
the parameter $\nu$ at the transition from pushed to pulled case are the
same as in the $n=1$ case, although we do not know how to obtain the
exact solution of the front profile in any $n>1$ case. In this sense,
all $n\geq 1$ models belong to the same universality class, as far as
the velocities and transition parameters are concerned. The numerical
calculations which we have performed in several $n\geq 1$ cases by the
shooting method are in excellent agreement with our analytical results
and support the assertions above.

We chose
$$\rho(u)=\kappa u(1-u),\eqn\vpeight$$
where $\kappa > 0$ is some adjustable parameter related to the decay rate of
the front profile. If we define $g(u)\equiv \rho'(u)+F(u)/\rho(u)$, then
$$g(u)=\kappa+\frac{1}{\kappa}+\frac{\nu}{\kappa}u^{n}-2\kappa u.\eqn\vpnine$$
Firstly, we consider the case $\nu > 0$ and $n>1$, and study the behavior of
the function
$\phi(u)=\nu u^{n}/\kappa-2\kappa u$ with $\kappa > 0$ fixed. It has an
extremum at the point $u_0=(2\kappa^2/{n\nu})^{1/{n-1}}$, which is determined
by
the condition
$\phi'(u_0)=n\nu u_{0}^{n-1}/\kappa-2\kappa=0$, and further, since $\phi''(u_0)
=n(n-1)\nu u_{0}^{n-2}/\kappa >0$,
 $\phi$ has a minimum at $u=u_0$. No
matter whether $u_0$ lies in or out of  $0\leq u\leq 1$, the supreme bound or
maximum is always reached at either of the two boundary sides, i.e. $u=0$ or
$1$. Thus, we have
$$G(\kappa)=\max_{0\leq u\leq 1}\{g(u)\}=\kappa+\frac{1}{\kappa}+\max\{0,
\frac{\nu}{\kappa}-2\kappa\}.\eqn\vpten$$
In the case of $n=1$ and $\nu>0$, we have $g(u)=\kappa+1/\kappa+(\nu/\kappa
-2\kappa)u$, so we have the same result \ \vpten \
as in the cases of $n>1$. There are two separate cases to consider below.
If $\nu/\kappa-2\kappa \geq 0$, i.e. $\kappa \leq \sqrt{\nu/2}$, then we have
$$\eqalign{
G(\kappa)=& \kappa+\frac{1}{\kappa}+(\frac{\nu}{\kappa}-2\kappa)\cr
	 =& \frac{1+\nu}{\kappa}-\kappa. \cr}\eqn\vpeleven$$
Since $G'(\kappa)=-{(1+\nu)}/\kappa^2-1 < 0$ for any $\kappa$, $G(\kappa)$ is a
monotonically decreasing function of $\kappa$, and the minimum is attained at
$\kappa=(\nu/2)^{1/2}$. Thus, in the interval $\nu \geq 2$, our estimate for
the minimal velocity is
$$c_0=\min_{\kappa>0}\{G(\kappa)\}=\frac{2+\nu}{\sqrt{2\nu}}.\eqn\vpthirteen$$
If $\nu/\kappa-2\kappa \leq 0$, i.e. $\kappa \geq \sqrt{\nu/2}$, then we
have
$$G(k)= \kappa+\frac{1}{\kappa}+0 \geq 2. \eqn\vpfourteen$$
with equality only when $\kappa=1$. Also, we must satisfy the condition
$1\geq \sqrt{\nu/2}$, i.e. $\nu \leq 2$. Thus, the minimum velocity is
$$c_0=\min_{\kappa>0}\{G(\kappa)\}=2.\eqn\vpfifteen$$
Secondly, we consider the case of $-1\leq \nu \leq 0$ and $n\geq 1$.
Because $\nu/\kappa
-2\kappa$ is always negative for any $\kappa >0$, we have the same results as
equation \vpfourteen \ and \vpfifteen. In summary, in all cases of $n\geq 1$,
we have the following minimal velocities
$$c_0=\cases{2, &for $-1\leq \nu\leq 2$ ;\cr
	     ({2+\nu}/{\sqrt{2\nu})}, &for $\nu\geq 2$ .\cr}
\eqn\vpsixteen$$

\newpage
{\bf \cl{B. Other Examples}}

A second more complicated example is the partial differential
equation
$$\frac{\pd u}{\pd t}=\frac{\pd^2 u}{\pd x^2} + u + d u^{n/2+1}- u^{n+1},\quad
n > 0.\eqn\vpseventeen$$
For $n=4$, it reduces to an equation considered by van Saarloos\rlap.\refmark
{\van} By using the
method of reduction of order of ODE, he obtained exact results for front
velocities $c_0$ and the transition parameter value $d_c$:
$$c_0=\cases{2 ,\quad &for $d\leq d_c={2}/{\sqrt{3}}$,\cr
	     ({-d+2\sqrt{d^2+4}}/{\sqrt{3})}, &for $d\geq d_c$.
\cr}\eqn\vpeighteen$$
The nonlinear source term in the equation \vpseventeen\ can be written as
$$F(u)=u(u_s^{n/2}-u^{n/2})(u_r^{n/2}+u^{n/2}),\eqn\vpnineteen$$
where $u_s^{n/2}=[d+(d^2+4)^{1/2}]/2$ and $u_r^{n/2}=[-d+(d^2+4)^{1/2}]/2$.
It always has an unstable steady state $u=0$ and an absolutely stable state
$u=u_s >0$. For simplicity, we only consider the propagation front connecting
the above two steady states.

In this case, we choose the trial function
$$\rho(u)=\kappa u(u_s^{n/2}-u^{n/2}),\quad \kappa>0, \quad 0\leq u\leq u_s,
\eqn\vptwenty$$
but {\it not} $\rho(u)=\kappa u(u_s-u)$ as in the former example.
As a result, we obtain the complete result
$$c_0=\cases{2 , &for $d\leq d_c\equiv{n}/{\sqrt{2(n+2)}}$,\cr
	     [{-nd+(n+2)\sqrt{d^2+4}]}/{2\sqrt{2(n+2)}}, &for
$d\geq d_c$.\cr}\eqn\vptwyfive$$
As a check, when $n=4$, the result \vpeighteen \ by van Saarloos\refmark{\van}
is recovered.

We remark that in these examples, only when $g(u)$ reaches its maximum at
$u=u_s>0$ with
$\kappa$ fixed is it possible for a transition to the pushed case to occur,
while
when $g(u)$ reaches its maximum at $u=0$ it is the pulled case
that occurs. We conjecture that this is typical. Based on this naive picture,
we have also studied several other
interesting cases:

\item{(1)} The Fisher-KPP equation\rlap,\refmark{\fisherkpp}
where $F(u)=u-u^n, n\geq 2$.
If we choose $\rho(u)=\kappa u(1-u^{n-1}), 0\leq u\leq 1$, we obtain
$c_0=2$.
\item{(2)} Schlogl's second model for chemical
reactions\rlap,\Ref\magyari{E. Magyari, {\sl J. Phys. A}
{\bf 15, L 139}(1982).}
where $F(u)=\gamma-\beta u +3 u^2
-u^3=(1-u)(u-u_1)(u-u_2), \beta=\gamma+2, 0<\gamma<1, u_1=1+(1-\gamma)^{1/2},
u_2=1-(1-\gamma)^{1/2}$.  We choose $\rho(u)=\kappa(u-1)(u_1-u), 1\leq u\leq
u_1$, and find that the exact result $c_0=3\sqrt{(1-\gamma)/2}$ is obtained.
\item{(3)} A generalised version of Fisher's model\rlap,\refmark{\hadeler}
where $F(u)=u(1-u)(u-\mu), 0\leq u\leq 1/2$.
If $\rho=\kappa u(1-u)$ is chosen, $c_0=1/\sqrt{2}-\sqrt{2}\mu$ is obtained.
This result is exactly the same as the perturbative RG result we obtained in
Section 4.

{\bf \chapter{Transition from Pushed to Pulled Cases}}

In this section, we consider how the transition point between pulled and pushed
cases is changed when a $p$-small perturbation is present.

Consider the following indicative example
$$
\frac{\pd u}{\pd t}=\frac{\pd^2 u}{\pd x^2} + u(u_s-u)(u_r+u+\epsilon u^n),
\quad n \ge 0 \eqn\done
$$
where $u_s=[d+\sqrt{d^2+4}]/2$, $u_r= [-d+\sqrt{d^2+4}]/2$, $d > 0$
and
$\epsilon$ is a perturbation parameter.  For $\epsilon = 0$ the
transition from the pulled to the pushed case occurs\refmark{\van}
as $d$ is increased from zero
at $d = d_{c} \equiv
1/\sqrt{2}$.   We can apply the perturbation approach only for
$d>d_{c}$, because we need the explicit formula for the unperturbed wave
front: $u_0(\xi)=u_s(1+Ae^{\kappa \xi})^{-1}$, where
$\kappa=[c_0+\sqrt{c_0^2-4}]/2$ and $c_0=[3\sqrt{d^2+4}-d]/2\sqrt{2}$, and
$A$ is a constant of integration.  For $d < d_c$, there is no known
formula for the unperturbed wavefront.

The zeroth
eigenfunction is $u'_0(\xi) \propto -\kappa u_0(1-u_0/u_s)$ and the
weight function $\rho(\xi)=\left( u_s/u_0-1\right)^{c_0/\kappa}$.
To determine the new transition point $d_c$, we
substitute into the formula \dcl\, and obtain, to $O(\epsilon)$,
$$
\delta c=\epsilon
\frac{u_s^{n+1}}{\kappa}\frac{\Gamma(n+4) \Gamma(2 - {c_0}/{\kappa})}
{3!\,\Gamma(n+2-{c_0}/{\kappa})}.
\eqn\dtwo
$$
We also have $\hat{c} = 2$ for $n>0$ and $\hat{c} = 2 \sqrt{1 + \epsilon u_{s}}
$ for $n = 0$.  Setting $c^*(d_c)=\hat c(d_c)$, we find that
$$
2\sqrt{1+\epsilon u_s(d_c)} = c_0(d_c) + \delta c(d_c).\eqn\setexact
$$
For $n=0$, we have $\delta c=\sqrt{2}\epsilon$
and
$$d_c=\frac{\sqrt{2 +\epsilon^2}+3\epsilon}{2}.\eqn\vpone$$
whilst for $n=1$, we have $\delta c= \epsilon/2 \sqrt{c_0^2-4}$ and
$$d_c=\frac{(1+3/2 \ \epsilon)}{\sqrt{2}}.\eqn\vptwo$$
For $n=2$, we have
$$\delta c=\epsilon/20\ u_s(c_0+3\sqrt{c_0^2-4})
(c_0-\sqrt{c_0^2-4})\sqrt{c_0^2-4}.\eqn\vpthree$$

We can also apply the variational methods to determine the velocity $c^*$
and transition point for any $n\geq 0$ and $d$. We choose
the trial function
$$\rho(u)=\kappa u(u_s-u),\quad 0\leq u\leq u_s,\quad \kappa>0.\eqn\vpfour$$
For $n=0 $, we find $d_c=[\sqrt{2+\epsilon^2}+3\epsilon]/2$
and
$$c=\cases{2\sqrt{1+u_s\epsilon}, &for $d\leq d_c$;\cr
	   c=c_0(d)+\sqrt{2}\epsilon, &for $d\geq d_c$.\cr}\eqn\vpfive$$
For $n=1$, we have the result $d_c=(1-\epsilon)/\sqrt{2(1+\epsilon)}$
and
$$c=\cases{2, &for $d\leq d_c$; \cr
    \sqrt{2/(1+\epsilon)}\sqrt{d^2+4}+ (\epsilon-1)/\sqrt{2(\epsilon+1)}
(d+\sqrt{d^2+4})/2, &for $d\geq d_c$.\cr}
\eqn\vpsix$$
If we expand the above
results to $O(\epsilon)$, we find that they coincide with the results
from perturbative
RG.  We suspect that the results from variational methods are
actually the exact ones for $n=0$ and $1$.

For $n=2$, we obtain
$$c=\cases{2, &for $d\leq d_c$,\cr
    \sqrt{2/(1+\epsilon)}(\sqrt{d^2 +4}+\epsilon u_s^2)-\sqrt{(1+\epsilon u_s)
/2}\ u_s, &for $d\geq d_c$,\cr}\eqn\vpseven$$
where $d_c$ is the positive root of the equation $\epsilon
u_s^3(d_c)+ u_s^2(d_c)-2=0$. If we expand the velocities in \vpseven\ to
$O(\epsilon)$,
we observe that these results are slightly greater than those from
perturbative RG, and yet give good upper bounds.  In fact, for $n\geq 2$
and small $\epsilon$, perturbative RG gives more accurate results on
velocities and transition points than the trial function we used.

{\bf \chapter{Concluding Remarks}}

In this paper, we have shown that for structurally stable fronts, a
renormalization group method can be used to compute the change in
the front speed when the governing equation is perturbed by a marginal
operator; further, by combining the structural stability principle with
RG, we are able to predict the uniquely selected front itself. Our
results apply to both the pulled and pushed cases.  We demonstrated that
the solvability condition widely used in studying pattern selection in
nonequilibrium systems is identical to the physical renormalizability
(observability) condition. We have also implemented a
variational principle which gives very good upper bounds and sometimes
exact results on front speeds, and which identifies the transition
between the pulled and pushed cases.

In future work, we will investigate the application of these methods for
systems where a spatial pattern forms behind the front.

\ACK
LYC would like to thank Yuhai Tu for introducing to him a useful paper on
dynamics of dislocations, and gratefully acknowledges support from the
National Science Foundation through grant number NSF-DMR-89-20538,
administered through the Illinois Materials Research Laboratory.  NDG
and YO gratefully acknowledge the partial support of the National
Science Foundation through grant number NSF-DMR-90-15791.

\newpage
\refout
\end